# Ultrasensitive Sub-monolayer Palladium Induced Chirality Switching and Topological Evolution of Skyrmions

Gong Chen[1,*], Colin Ophus[2], Roberto Lo Conte[3,4], Roland Wiesendanger[4], Gen Yin,[1] Andreas K. Schmid,[2] and Kai Liu[1,*]

[1]Department of Physics, Georgetown University, Washington, D.C. 20057, United States

[2]NCEM, Molecular Foundry, Lawrence Berkeley National Laboratory, Berkeley, California, 94720, United States

[3]Department of Materials Science & Engineering, University of California, Berkeley, California 95720, United States

[4]Department of Physics, University of Hamburg, D-20355 Hamburg, Germany

**ABSTRACT:** Chiral spin textures are fundamentally interesting, with promise for device applications. Stabilizing chirality is conventionally achieved by introducing Dzyaloshinskii–Moriya interaction (DMI) in asymmetric multilayers where the thickness of each layer is at least a few monolayers. Here we report an ultrasensitive chirality switching in (Ni/Co)$_n$ multilayer induced by capping with only 0.22 monolayer of Pd. Using spin-polarized low-energy electron microscopy, we monitor the gradual evolution of domain walls from left-handed to right-handed Néel walls and quantify the DMI induced by the Pd capping layer. We also observe the chiral evolution of a skyrmion during the DMI switching, where no significant topological protection is found as the skyrmion winding number varies. This corresponds to a minimum energy cost of < 1 attojoule during the skyrmion chirality switching. Our results demonstrate the detailed chirality evolution within skyrmions during the DMI sign switching, which is relevant to practical applications of skyrmionic devices.

**KEYWORDS:** magnetic skyrmions, chiral spin textures, Dzyaloshinskii-Moriya interaction, topological protection, spintronics

**Corresponding Authors**

*Email: gchenncem@gmail.com (G.C.)

*Email: Kai.Liu@georgetown.edu (K.L.)

Magnetic skyrmions are bubble-like topological spin textures, characterized by a topological charge (or skyrmion number)[1-4]. Due to their topologically protected spin configurations, magnetic skyrmions have the potential to be used as low-dissipation information carriers in spintronic applications, such as skyrmion-based memory[5, 6], logic devices[7], artificial neurons,[8] or random number generators.[9] They may also find applications in more complex device architectures such as three-dimensional (3D) racetrack memories[10] or interconnected networks[11].

One of the main mechanisms to stabilize skyrmions is the Dzyaloshinskii–Moriya interaction (DMI)[12, 13]. DMI usually occurs under conditions where the inversion symmetry is broken, e.g., in the B20-like group of bulk compounds[14, 15] or in thin films[16]. The real-space *chiral* character of skyrmions was directly imaged in bulk FeCoSi with B20 structures using Lorentz transmission electron microscopy,[17] and in Fe/Ir [18] or Pd/Fe/Ir [19] thin films using spin-polarized scanning tunneling microscopy. Later, chiral skyrmion structures were observed by other imaging techniques such as spin-polarized low energy electron microscopy (SPLEEM) [20, 21], photoemission electron microscopy [22], or indirectly through their dynamic characteristics [23]. Skyrmions can also be stabilized in artificial structures, where the chirality is achieved through the asymmetric dot-shape [24, 25].

In perpendicularly magnetized systems that host skyrmions, the DMI stabilizes the spin structure of a skyrmion as homochiral, where the chirality along the in-plane boundary is fixed. Such homochirality in a skyrmion gives rise to its topological properties [26], manifested in, e.g., the skyrmion Hall effect [27, 28] or the topological protection [1, 29]. The topological protection of skyrmions, e.g. its stability against thermal effect or magnetic field that may trigger the skyrmion annihilation, appears to be entropy-limited [30] and with longer lifetime compared to topologically trivial bubbles [31]. In certain confined nanostructures, topological protection has been found to be insufficient because of possible annihilation at the edges [32].

Recently, tuning of the DMI has been demonstrated by a variety of methods, such as controlling the film thickness of DMI materials [33, 34], electrical field [35], ionic migration [36], chemisorption[37, 38], or ion irradiation [39]. It represents a potentially effective means to tailor the skyrmion stability. However, whether a skyrmion structure may still maintain fixed chirality in a weak DMI system has been rarely explored experimentally. For example, observing the evolution of the detailed spin structure within a skyrmion while the DMI is gradually changing its sign may provide insight on the stability of the topological structure of a skyrmion.

Here we report an extremely sensitive fine-tuning of the chirality across the zero-DMI point in (Ni/Co)$_n$ multilayers, via a sub-monolayer Pd capping layer [thickness $d_{cap}$ up to 0.22 monolayer in total; here monolayer (ML) refers to a single layer in (111) stacking]. We monitor the domain wall's magnetization angle evolution during the transition, which shows a linear dependence as a function of $d_{cap}$. The DMI induced by the Pd capping layer is quantified as ~0.012 meV/atom for every 0.02 ML $d_{cap}$ change, or effectively 0.6 meV/atom per ML Pd. This precise control of the DMI has allowed the observation of microscopic spin structures of a skyrmion bubble, and its evolution as a function of $d_{cap}$.

We find a $d_{cap}$-dependent transition from left-handed to mixed chirality and then to right-handed chirality within the skyrmion bubble. This transition demonstrates an abrupt change of topological charge during the DMI switching, where a bubble with mixed chirality appears when the magnitude of the DMI is less than ~0.02 meV.

The nature of the interfacial DMI, e.g., its sign and magnitude, is obtained via the direct observation of magnetic chirality. Imaging magnetic chirality has been one of the major approaches to understand the interfacial DMI [26]. Pioneering observations of cycloidal spin spirals using spin-polarized scanning tunneling microscopy revealed the role of the interfacial DMI in ultrathin films [16, 18, 40-42], e.g., the orientation of the DMI vector, $D_{ij}$, and its magnitude with respect to the exchange interaction can be quantified. More recent observations of chirality in domain walls, particularly the thickness-driven transition of chiral Néel walls to achiral Bloch walls also allows the quantification of the interfacial DMI [33, 43, 44].

To explore the sub-monolayer metal induced DMI as well as the skyrmion spin structure near the DMI switching region, we adapt a tunable DMI system where the DMI induced at the buried interface can be controlled via the competition between two opposite DMI materials, e.g., an Ir layer on a Pt(111) crystal [33] or a Pd layer on a W(110) crystal [37]. The advantage of this approach for quantifying weak DMI variation is the possibility of fine-tuning of the DMI at the buried interface; i.e., both magnitude and sign of the DMI can be precisely controlled right around the zero-DMI point via the thickness of the inserted Pd layer (to be distinguished from the Pd capping layer thickness $d_{cap}$). The system studied here is a (Ni/Co)$_n$ multilayer on Pd/W(110). Figure 1 shows the $d_{Pd}$-dependent evolution of magnetic chirality in Ni/1 ML Co/2 ML Ni/1 ML Co/Pd ($d_{Pd}$ varies)/W(110), and the chirality is derived from the compound SPLEEM image (Figure 1a-d). Perpendicularly magnetized domains with right-handed chiral domain walls are observed at $d_{Pd}$ = 1.9 ML (Figure 1a); i.e., domain wall magnetization is pointing from the +M$_z$ domain to the -M$_z$ domain (↑→↓←↑), and left-handed chiral domain walls are observed at $d_{Pd}$ =2.0 ML (Figure 1c), i.e. domain wall magnetization is pointing from the -M$_z$ domain to the +M$_z$ domain (↓→↑←↓). A statistical survey of the magnetic chirality within domain walls is presented in the $\alpha$-histogram (Figure 1b,d), where $\alpha$ is the angle between the domain boundary's normal direction (vector **n**) and the magnetization direction within the domain wall (vector **m**). A single peak at 180° in Figure 1b and at 0° in Figure 1d further support the chirality observation in SPLEEM images, indicating right-handed and left-handed Néel walls, respectively. Figure 1e shows the quantified chirality as a function of $d_{Pd}$ [33, 38], revealing a sharp transition between $d_{Pd}$ =1.9 ML and 2.0 ML. This $d_{Pd}$ dependent trend is consistent with the previous observation of chirality in the Ni/Co/Pd/W(110) system [38], where the chirality at the thinner Pd region is dominated by W-like DMI , favoring right-handed DMI, and that at the thicker Pd region is dominated by Pd-like DMI, favoring left-handed DMI. The thickness of top Ni layers is slightly tuned to optimize the magnetic anisotropy for domain wall imaging, and the Ni thickness variation here shows no effect on the effective DMI[37].

The tunable DMI system allows the quantification of the DMI induced by the additional layer on the top surface [37, 38]. It also allows the assessment of the role of the Pd capping layer on the net DMI, with sub-monolayer sensitivity. We have systematically studied the domain wall evolution induced by the sub-monolayer Pd capping layer by first varying the Pd thickness in the middle of the Pd(capping)/(Ni/Co)$_n$/Pd($d_{Pd}$)/W(110) stack, $d_{Pd}$, and we found no appreciable chirality switching on either the right-handed/W-like DMI side or the left-handed/Pd-like DMI side when $d_{Pd}$ is far away from the chirality transition thickness. However, in the sample with very weak Pd-like DMI where $d_{Pd}$=2.05 ML is very close to the 0-DMI point, a gradual and complete transition of the chirality is observed within a range of 0.22 ML Pd capping layer. Figure 2a shows the detailed evolution of domain structures at the same place on the sample surface, where the arrows represent the magnetization of domain walls, highlighting a gradual transition of the domain wall type/chirality as a function of Pd capping layer thickness $d_{cap}$. Here no significant pinning effect is observed in Fig. 2a where the domain evolves due to the $d_{cap}$-induced anisotropy change. Note that the usual presence of a uniaxial in-plane anisotropy due to the uniaxial strain from the growth of fcc-Pd on bcc-W(110) is eliminated by fine-tuning the thickness ratio between Co and Ni layer [37], because of their opposite magneto-elastic constants [45].

To be more quantitative, the $\alpha$–histogram is used to illustrate the statistics of the domain wall angle distribution presented in Fig. 2a, as shown in Fig. 2b. A gradual transition of peaks is observed, from a single peak at 0° for $d_{cap}$ = 0ML, to double peaks at ∼ ±90° for $d_{cap}$ =0.14ML, and to a single peak at 180° for $d_{cap}$ = 0.22ML. The peak position is indicated by blue arrows, showing an essentially linear evolution of the wall angle with $d_{cap}$, as summarized in Fig. 2c. A full 180° domain wall angle switching, along with a complete chirality switching from left-handed to right-handed Néel wall, is clearly achieved by growing a 0.22ML Pd capping layer.

The direct observation of spin structures may allow the quantification of the DMI. The observation of domain walls using SPLEEM provides information on their type and chirality[33, 43], e.g., chiral Néel walls in the thinner film where the DMI dominates and achiral Bloch walls in the thicker film where the dipolar energy dominates. Therefore, one can estimate the strength of the DMI by bracketing the range of the DMI if a transition of chiral Néel wall to achiral Bloch wall is observed [43]. Alternatively, in other systems the DMI can be quantified by comparing the DMI to the calculated dipolar energy when the wall angle is pointing at 45° to the domain boundary direction[37, 44], where the 45° wall is a result of the competition between the DMI and dipolar energy.

Following this approach, the sub-monolayer Pd induced DMI at the top interface can be estimated via the systematic observation of the domain wall angle in Fig. 2c. For instance, the 45° wall between left-handed Néel type and achiral Bloch type is observed at a Pd capping layer thickness $d_{cap}$≈0.06 ML, indicating that the dipolar energy and the DMI energy favoring left-handed chirality are comparable. The same applies to the 135° wall between right-handed Néel type and achiral Bloch type where dipolar energy and the DMI energy favoring right-handed chirality are comparable at $d_{cap} \approx 0.19$ ML, because the averaged wall angle at $d_{cap}=0.18$ ML is slightly below 135° and that at $d_{cap}=0.20$ ML is slightly above 135°. By calculating the dipolar energy contribution in a (Ni/Co)$_2$ film[33], we could indirectly estimate the DMI energy difference between $d_{cap} \approx 0.06$ ML and 0.19 ML (mentioned above) as twice the dipolar contribution from (Ni/Co)$_2$, which is 0.077 meV/atom, or effectively 0.6 meV/atom for a full monolayer of Pd cap on top of (Ni/Co)$_2$. This is also equivalent to an increase of the DMI by 0.012 meV/atom for every 0.02 ML Pd, which will be used to understand the skyrmion evolution in the next section. Note that this number also agrees with the 0.41 meV/atom DMI induced by 1ML Pd thickness change in the Ni/Co/Pd/W(110) system[37].

The quantitative sub-monolayer DMI tuning provides a platform to understand the evolution of the skyrmion spin texture in a DMI switching system in detail. The skyrmion structure is explored in the same $d_{cap}$ dependent measurements in Pd(capping)/(Ni/Co)$_n$/Pd(2.05 ML)/W(110) system. Figure 3a shows the sequence of the compound SPLEEM image of a skyrmion bubble, where the orientation of the domain wall structure is highlighted by the white arrows. The slightly irregular shape of this skyrmion bubble might be related to the morphology-induced variation of magnetic interactions. In the case of the skyrmion bubble at $d_{cap}=0$ ML, all of the in-plane magnetization components are pointing from the gray domain (-M$_z$) to the black domain (+M$_z$), showing that this is a typical left-handed hedgehog (Néel-type) skyrmion.

With increasing Pd capping thickness, the spin structure along the boundary is gradually tilting from the Néel type ($d_{cap}=0$ ML) to Bloch type ($d_{cap}=0.08$ ML). Simultaneously the magnetic chirality evolves from a homochiral Néel wall at $d_{cap}=0$ ML to an achiral Bloch wall, where the Bloch component with both left- and right-handed Bloch chirality (arrows pointing to the left/right of the local domain wall normal) is evident in sections of the boundary at $d_{cap}=0.08$ ML in Figure 3a. At $d_{cap}=0.12$, 0.14 and 0.16 ML, interestingly both left-handed (arrows pointing radially inward for +M$_z$ bubble) and right-handed Néel components (arrows pointing radially outward) appear in a single skyrmion bubble, as shown in Figure 3a. This mixed Néel chirality further evolves to a fully right-handed Néel wall at $d_{cap}=0.18$, 0.20 and 0.22 ML. We attribute the gradual diameter variation of skyrmion bubble in Fig. 3a to the anisotropy induced by a Pd capping layer, rather than the DMI dependence of the skyrmion diameter [35, 46]. The latter effect is expected to be non-monotonic as the DMI goes across zero, and its absence in this study is possibly due to the smaller magnitude of the DMI.

A skyrmion is usually considered to be topologically protected[1], and its skyrmion winding number[2, 47], i.e., how many times the magnetization vector is wrapped around a unit sphere, would remain the same until the skyrmion is annihilated[19, 48, 49]. In the case of varying DMI, the transition of homochiral Néel type to homochiral Bloch type was observed in skyrmions with a diameter less than 200 nm[37]. Such a transition is considered topologically smooth; i.e., it does not change the skyrmion winding number. However, the emergence of the mixed chirality of the Néel component while the DMI is varied, as seen in Figure 3b-d, shows that the topology of this skyrmion is not protected. In fact, from these images one can quantify the skyrmion winding number by tracing the skyrmion boundary magnetization all the way around the skyrmion. Figure 3f shows the $d_{cap}$-dependent accumulative magnetization rotation around the skyrmion boundary. In the $d_{cap}$= 0ML case, magnetization rotates gradually by 2π, which is also obvious in Figure 3a, corresponding to a winding number of 1. This 2π rotation is conserved in the subsequent images up to the $d_{cap}$=0.12 ML image. However, for $d_{cap}$=0.14 and 0.16 ML the accumulative magnetization rotation is 0, indicating topologically trivial bubbles (Figure 3e). In subsequent images for $d_{cap}$=0.18 – 0.22 ML the accumulative magnetization rotation is back to 2π (winding number 1). The $d_{cap}$-dependent skyrmion number is summarized in Figure 3g.

Neighboring domain wall segments with opposite Néel components may induce Bloch lines, highlighted by yellow arrows in Figure 3c as an example (ones with insufficient/gradual magnetization rotation are not counted). The number of Bloch lines as a function of $d_{cap}$ as well as the quantified DMI strength is shown in Figure 3g. It is interesting to point out that the phase with mixed Néel-components only appears when the DMI is close to zero, near $d_{cap}$ ~ 0.14 ML, where the threshold DMI sufficient to stabilize the chirality could be estimated as $0.020 \pm 0.006$ meV/atom for the left-handed side (between $d_{cap}$ = 0.10 and 0.12 ML), and $-0.015 \pm 0.006$ meV/atom for the right-handed side (between $d_{cap}$ = 0.16 and 0.18 ML). The total DMI energy in such skyrmion bubbles can be approximated as -$DLt\pi$, where $D$ is the DMI constant, $t$ is the film thickness, $L$ is the perimeter of the skyrmion bubble, and π corresponds to the 180° rotation within the skyrmion boundary [33]. Therefore, the DMI energy variation $\Delta E_{DMI}$, from the left-handed threshold to the right-handed counterpart, can be estimated as $0.93 \times 10^{-18}$ J with $\Delta D \sim 0.035$ meV/atom (which is 0.085 mJ/m$^2$), skyrmion diameter ~1.2 µm and film thickness $t$=0.92 nm. Note that this DMI energy variation may vary with the skyrmion diameter, e.g. $\Delta E_{DMI} \approx 0.78 \times 10^{-19}$ J for a diameter of ~100 nm, which reasonably agrees with the micromagnetic study in Ref. [50]. This estimation suggests that the minimum energy cost to switch the handedness of a skyrmion with the DMI and the film thickness comparable to this study would be on the order of $10^{-19}$ J (with 100 nm skyrmion diameter) or $10^{-18}$ J (with 1 µm diameter).

This sub-attojoule energy cost associated with skyrmion chirality switching, or "writing", near the zero-DMI point is attractive for energy-efficient skyrmionics devices. In particular, DMI-dependent domain wall type may strongly affect the current-driven dynamics of skyrmions, e.g., chirality-dependent propagation direction[51]. Moreover, the DMI-dependent variation of skyrmion number observed in this work might be relevant to the control of the skyrmion Hall angle[27]. Once the skyrmion chirality is switched (written), the energy barrier may be enlarged to ensure thermal stability through stimuli such as magnetic field,[1-3] current,[19] gating,[48] temperature[52, 53], or chemisorption[49]. That is, the energetics for the writing process can be separated from thermal stability, analogous to that in heat-assisted magnetic recording (HAMR) where writing is done at elevated temperatures with much reduced anisotropy and thermal stability is anchored at room temperature when anisotropy is much larger.[54]

In summary, our results demonstrate an extremely sensitive and complete switching of the DMI and domain wall chirality induced by a sub-monolayer Pd capping layer of just 0.22 ML. The linear and gradual transition of the domain wall type with Pd cap thickness allows the quantification of the induced DMI on the surface of [Ni/Co]$_2$, which is effectively 0.6 meV/atom for 1 ML $d_{cap}$. We have further observed the spin structure evolution of a skyrmion bubble during the DMI switching, and found that mixed Néel-components with different skyrmion winding numbers appear when the strength of the DMI was smaller than $\sim|0.015 - 0.020|$ meV/atom. The minimum DMI energy barrier for the chirality switching of a skyrmion is estimated to range from $10^{-19}$ (100nm skyrmion diameter) to $10^{-18}$ $J$ (1μm diameter). Our results show evidence for the instability of the skyrmion topological protection in the zero DMI limit and demonstrate a sensitive and energy-efficient DMI handle to control the chiral nature of magnetic skyrmions.

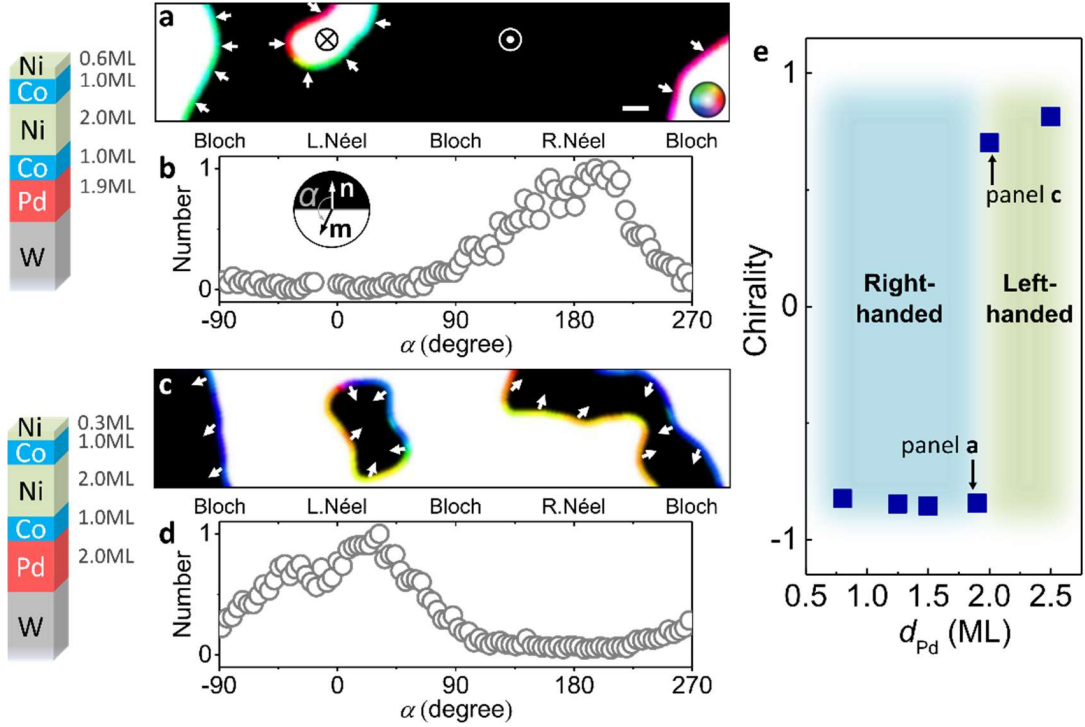

**Figure 1.** Tunable DMI system of (Ni/Co)$_2$/Pd($d_{Pd}$)/W(110) with perpendicular magnetization. (a) Compound SPLEEM image at $d_{Pd}$ =1.9 ML. Black/white regions represent up/down domains (see symbols), and the color wheel shows the in-plane magnetization direction within domain walls (also highlighted by white arrows). Scale bar is 200 nm. (b) Histogram of angle $\alpha$ between boundary normal vector **n** and in-plane magnetization **m** (see inset), measured pixel-by-pixel along domain boundaries, showing right-handed Néel-type chirality. (c) Compound SPLEEM image at $d_{Pd}$ =2.0 ML. (d) Histogram of angle $\alpha$, showing left-handed Néel-type chirality. (e) Chirality evolution as a function of $d_{Pd}$. Chirality is defined as ($N_{Left}$-$N_{Right}$)/ ($N_{Left}$+$N_{Right}$), $N_{Left}$ is the number of domain wall pixels where $\alpha$ is between -90° and 90°, and $N_{Right}$ is the number of domain wall pixels where $\alpha$ is between 90° and 270°.

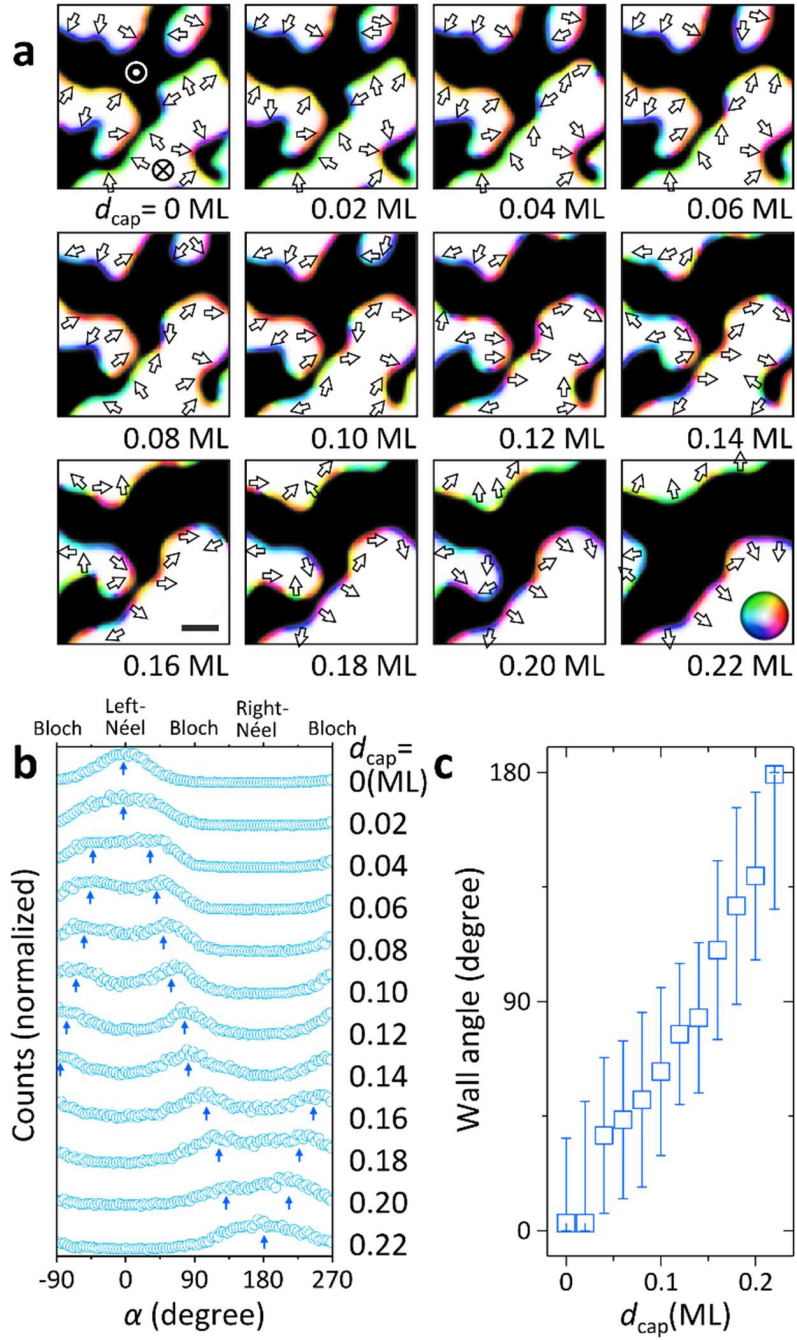

**Figure 2.** Chirality evolution of the Pd(capping)/(Ni/Co)$_2$/Pd(2.05 ML)/W(110) as a function of Pd capping layer thickness $d_{cap}$. (a) SPLEEM image sequence at varying Pd capping thickness $d_{cap}$. Black/white regions represent up/down domains, and the color wheel highlights the in-plane magnetization direction within domain walls (also shown by white arrows). Scale bar is 500 nm. (b) $d_{cap}$-dependent $\alpha$-histogram, indicating a gradual shift of the peaks (blue arrows). (c) Averaged domain wall angle with respect to the boundary's normal direction (from white to black) as a function of $d_{cap}$, showing a gradual transition.

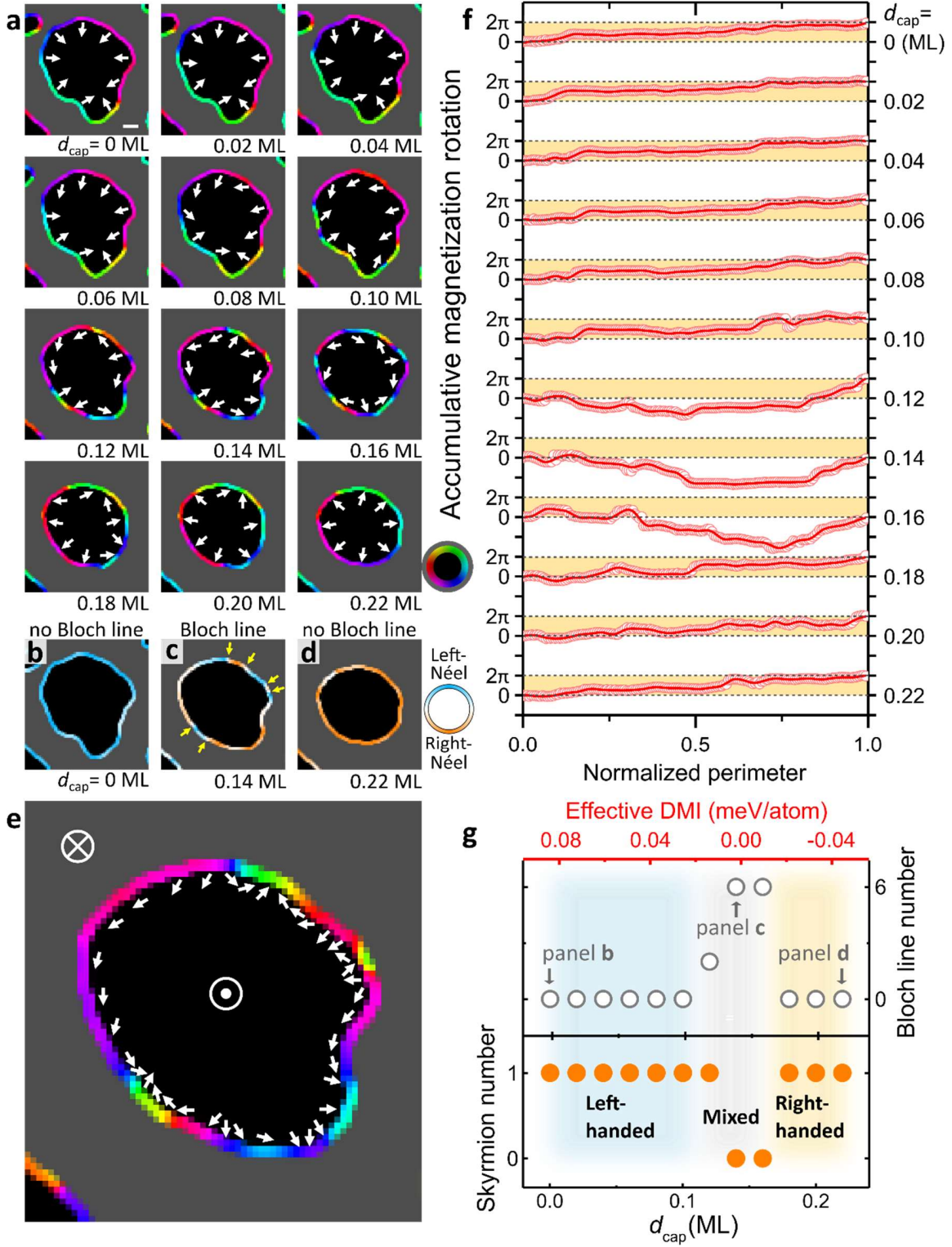

**Figure 3**. DMI-dependent evolution of skyrmion structures in (Ni/Co)$_2$/Pd(2.05 ML)/W(110) capped by Pd. (a) Sequence of compound SPLEEM images with various Pd $d_{cap}$. Black/gray indicates up/down domains, and the color wheel highlights the in-plane magnetization direction on the skyrmion boundary (also shown by the white arrows) .

Scale bar is 200 nm. (b-d) Decomposed Néel-component images of the same area at $d_{cap}$ = 0, 0.14 and 0.22 ML. The blue, white, and orange color wheel highlights left-handed Néel-type, Bloch-type, and right-handed Néel-type domain wall segments, respectively. Yellow arrows in panel c indicate the positions of Bloch lines, where the Néel-type chirality are opposite on both sides. (e) Enlarged compound SPLEEM image for $d_{cap}$ = 0.14 ML in panel a, highlighting the detailed magnetization rotation along the boundary (by white arrows), where the accumulative magnetization rotation is 0. (f) $d_{cap}$-dependent accumulative magnetization rotation around the perimeter of the skyrmion bubble, the counting starts from the bottom of the image with counterclockwise direction. The color-bar highlights the range of magnetization rotation for homochiral cases, where the rotation is expected to be monotonic between 0 and $2\pi$. (g) Evolution of Bloch line number and skyrmion number as a function of $d_{cap}$ (lower x-axis) and the effective DMI strength (upper x-axis). The Bloch line number represents the frequency of Néel-component switching within one skyrmion bubble, where the number 0 means fully chiral and a finite number indicates a mixed chirality.

AUTHOR INFORMATION

**ORCID**

Gong Chen: 0000-0002-5102-7487

Colin Ophus: 0000-0003-2348-8558

Roberto Lo Conte: 0000-0002-5050-9978

Roland Wiesendanger: 0000-0002-0472-4183

Gen Yin: 0000-0001-5827-5101

Andreas K. Schmid: 0000-0003-0035-3095

Kai Liu: 0000-0001-9413-6782

**Author Contributions**

G.C., A.K.S. and K.L. conceived the study. G.C. performed SPLEEM experiments. G.C., R.L.C., and C.O. analyzed SPLEEM data. G.C., R.L.C, R.W, G.Y., A.K.S. and K.L discussed and interpreted the results. G.C., A.K.S. and K.L. prepared the manuscript. All authors contributed to the manuscript revision.

**Acknowledgment**

We thank Tianping Ma for insightful discussions. This work has been supported in part by the NSF (Grant No. DMR-2005108) and SMART (Grant No. 2018-NE-2861), one of seven centers of nCORE, a Semiconductor Research Corp. program, sponsored by the National Institute of Standards and Technology (NIST). Work at the Molecular Foundry was supported by the Office of Science, Office of Basic Energy Sciences, of the U.S. Department of Energy under Contract No. DE-AC02-05CH11231. C.O. acknowledges support from the U.S. Department of Energy Early Career Research Program. R.L.C. and R.W. acknowledge financial support by the European Union via an International Marie Skłodowska-Curie Fellowship (Grant No. 748006 - SKDWONTRACK).

**REFERENCES**

1. Fert, A.; Cros, V.; Sampaio, J. Skyrmions on the track. *Nat. Nanotechnol.* **2013,** 8, (3), 152-156.
2. Nagaosa, N.; Tokura, Y. Topological properties and dynamics of magnetic skyrmions. *Nat. Nanotechnol.* **2013,** 8, (12), 899-911.
3. Muhlbauer, S.; Binz, B.; Jonietz, F.; Pfleiderer, C.; Rosch, A.; Neubauer, A.; Georgii, R.; Boni, P. Skyrmion Lattice in a Chiral Magnet. *Science* **2009,** 323, (5916), 915-919.
4. Rossler, U. K.; Bogdanov, A. N.; Pfleiderer, C. Spontaneous skyrmion ground states in magnetic metals. *Nature* **2006,** 442, (7104), 797-801.
5. Iwasaki, J.; Mochizuki, M.; Nagaosa, N. Current-induced skyrmion dynamics in constricted geometries. *Nat. Nanotechnol.* **2013,** 8, (10), 742-747.
6. Sampaio, J.; Cros, V.; Rohart, S.; Thiaville, A.; Fert, A. Nucleation, stability and current-induced motion of isolated magnetic skyrmions in nanostructures. *Nat. Nanotechnol.* **2013,** 8, (11), 839-844.
7. Zhang, X. C.; Ezawa, M.; Zhou, Y. Magnetic skyrmion logic gates: conversion, duplication and merging of skyrmions. *Sci. Rep.* **2015,** 5, 9400.
8. Song, K. M.; Jeong, J.-S.; Pan, B.; Zhang, X.; Xia, J.; Cha, S.; Park, T.-E.; Kim, K.; Finizio, S.; Raabe, J.; Chang, J.; Zhou, Y.; Zhao, W.; Kang, W.; Ju, H.; Woo, S. Skyrmion-based artificial synapses for neuromorphic computing. *Nat. Electron.* **2020,** 3, (3), 148-155.
9. Wang, K.; Zhang, Y.; Bheemarasetty, V.; Zhou, S.; Ying, S.-C.; Xiao, G. Single skyrmion true random number generator using local dynamics and interaction between skyrmions. *Nat. Commun.* **2022,** 13, (1), 722.
10. Parkin, S.; Yang, S. H. Memory on the racetrack. *Nat. Nanotechnol.* **2015,** 10, (3), 195-198.
11. Burks, E. C.; Gilbert, D. A.; Murray, P. D.; Flores, C.; Felter, T. E.; Charnvanichborikarn, S.; Kucheyev, S. O.; Colvin, J. D.; Yin, G.; Liu, K. 3D Nanomagnetism in Low Density Interconnected Nanowire Networks. *Nano Lett.* **2021,** 21, 716−722.
12. Dzyaloshinsky, I. A Thermodynamic Theory of Weak Ferromagnetism of Antiferromagnetics. *J. Phys. Chem. Solids* **1958,** 4, (4), 241-255.
13. Moriya, T. Anisotropic Superexchange Interaction and Weak Ferromagnetism. *Phys. Rev.* **1960,** 120, (1), 91-98.
14. Uchida, M.; Onose, Y.; Matsui, Y.; Tokura, Y. Real-space observation of helical spin order. *Science* **2006,** 311, (5759), 359-361.
15. Huang, S. X.; Chien, C. L. Extended Skyrmion phase in epitaxial FeGe(111) thin films. *Phys. Rev. Lett.* **2012,** 108, (26), 267201.
16. Bode, M.; Heide, M.; von Bergmann, K.; Ferriani, P.; Heinze, S.; Bihlmayer, G.; Kubetzka, A.; Pietzsch, O.; Blugel, S.; Wiesendanger, R. Chiral magnetic order at surfaces driven by inversion asymmetry. *Nature* **2007,** 447, (7141), 190-193.
17. Yu, X. Z.; Onose, Y.; Kanazawa, N.; Park, J. H.; Han, J. H.; Matsui, Y.; Nagaosa, N.; Tokura, Y. Real-space observation of a two-dimensional skyrmion crystal. *Nature* **2010,** 465, (7300), 901-904.
18. Heinze, S.; von Bergmann, K.; Menzel, M.; Brede, J.; Kubetzka, A.; Wiesendanger, R.; Bihlmayer, G.; Blügel, S. Spontaneous atomic-scale magnetic skyrmion lattice in two dimensions. *Nat. Phys.* **2011,** 7, (9), 713-718.

19. Romming, N.; Hanneken, C.; Menzel, M.; Bickel, J. E.; Wolter, B.; von Bergmann, K.; Kubetzka, A.; Wiesendanger, R. Writing and deleting single magnetic skyrmions. *Science* **2013,** 341, (6146), 636-639.

20. Chen, G.; Mascaraque, A.; N'Diaye, A. T.; Schmid, A. K. Room temperature skyrmion ground state stabilized through interlayer exchange coupling. *Appl. Phys. Lett.* **2015,** 106, (24), 242404.

21. Lo Conte, R.; Nandy, A. K.; Chen, G.; Cauduro, A. L. F.; Maity, A.; Ophus, C.; Chen, Z. J.; N'Diaye, A. T.; Liu, K.; Schmid, A. K.; Wiesendanger, R. Tuning the Properties of Zero-Field Room Temperature Ferromagnetic Skyrmions by Interlayer Exchange Coupling. *Nano Lett.* **2020,** 20, (7), 4739-4747.

22. Boulle, O.; Vogel, J.; Yang, H.; Pizzini, S.; de Souza Chaves, D.; Locatelli, A.; Mentes, T. O.; Sala, A.; Buda-Prejbeanu, L. D.; Klein, O.; Belmeguenai, M.; Roussigne, Y.; Stashkevich, A.; Cherif, S. M.; Aballe, L.; Foerster, M.; Chshiev, M.; Auffret, S.; Miron, I. M.; Gaudin, G. Room-temperature chiral magnetic skyrmions in ultrathin magnetic nanostructures. *Nat. Nanotechnol.* **2016,** 11, (5), 449-454.

23. Jiang, W. J.; Upadhyaya, P.; Zhang, W.; Yu, G. Q.; Jungfleisch, M. B.; Fradin, F. Y.; Pearson, J. E.; Tserkovnyak, Y.; Wang, K. L.; Heinonen, O.; te Velthuis, S. G. E.; Hoffmann, A. Blowing magnetic skyrmion bubbles. *Science* **2015,** 349, (6245), 283-286.

24. Sun, L.; Cao, R. X.; Miao, B. F.; Feng, Z.; You, B.; Wu, D.; Zhang, W.; Hu, A.; Ding, H. F. Creating an artificial two-dimensional Skyrmion crystal by nanopatterning. *Phys. Rev. Lett.* **2013,** 110, (16), 167201.

25. Gilbert, D. A.; Maranville, B. B.; Balk, A. L.; Kirby, B. J.; Fischer, P.; Pierce, D. T.; Unguris, J.; Borchers, J. A.; Liu, K. Realization of ground-state artificial skyrmion lattices at room temperature. *Nat. Commun.* **2015,** 6, 8462.

26. Jiang, W.; Chen, G.; Liu, K.; Zang, J.; te Velthuis, S. G. E.; Hoffmann, A. Skyrmions in magnetic multilayers. *Phys. Rep.* **2017,** 704, 1-49.

27. Jiang, W.; Zhang, X.; Yu, G.; Zhang, W.; Wang, X.; Benjamin Jungfleisch, M.; Pearson, John E.; Cheng, X.; Heinonen, O.; Wang, K. L.; Zhou, Y.; Hoffmann, A.; te Velthuis, Suzanne G. E. Direct observation of the skyrmion Hall effect. *Nat. Phys.* **2016,** 13, (2), 162-169.

28. Litzius, K.; Lemesh, I.; Kruger, B.; Bassirian, P.; Caretta, L.; Richter, K.; Buttner, F.; Sato, K.; Tretiakov, O. A.; Forster, J.; Reeve, R. M.; Weigand, M.; Bykova, L.; Stoll, H.; Schutz, G.; Beach, G. S. D.; Klaui, M. Skyrmion Hall effect revealed by direct time-resolved X-ray microscopy. *Nat. Phys.* **2017,** 13, (2), 170-175.

29. Yin, G.; Li, Y. F.; Kong, L. Y.; Lake, R. K.; Chien, C. L.; Zang, J. D. Topological charge analysis of ultrafast single skyrmion creation. *Phys. Rev. B* **2016,** 93, (17), 174403.

30. Wild, J.; Meier, T. N. G.; Pollath, S.; Kronseder, M.; Bauer, A.; Chacon, A.; Halder, M.; Schowalter, M.; Rosenauer, A.; Zweck, J.; Muller, J.; Rosch, A.; Pfleiderer, C.; Back, C. H. Entropy-limited topological protection of skyrmions. *Sci. Adv.* **2017,** 3, (9), e1701704.

31. Je, S. G.; Han, H. S.; Kim, S. K.; Montoya, S. A.; Chao, W. L.; Hong, I. S.; Fullerton, E. E.; Lee, K. S.; Lee, K. J.; Im, M. Y.; Hong, J. I. Direct Demonstration of Topological Stability of Magnetic Skyrmions via Topology Manipulation. *Acs Nano* **2020,** 14, (3), 3251-3258.

32. Cortes-Ortuno, D.; Wang, W. W.; Beg, M.; Pepper, R. A.; Bisotti, M. A.; Carey, R.; Vousden, M.; Kluyver, T.; Hovorka, O.; Fangohr, H. Thermal stability and topological protection of skyrmions in nanotracks. *Sci. Rep.* **2017,** 7, 4060.

33. Chen, G.; Ma, T.; N'Diaye, A. T.; Kwon, H.; Won, C.; Wu, Y.; Schmid, A. K. Tailoring the chirality of

magnetic domain walls by interface engineering. *Nat. Commun.* **2013,** 4, 2671.

34. Hrabec, A.; Porter, N. A.; Wells, A.; Benitez, M. J.; Burnell, G.; McVitie, S.; McGrouther, D.; Moore, T. A.; Marrows, C. H. Measuring and tailoring the Dzyaloshinskii-Moriya interaction in perpendicularly magnetized thin films. *Phys. Rev. B* **2014,** 90, (2), 020402(R).

35. Srivastava, T.; Schott, M.; Juge, R.; Krizakova, V.; Belmeguenai, M.; Roussigne, Y.; Bernand-Mantel, A.; Ranno, L.; Pizzini, S.; Cherif, S. M.; Stashkevich, A.; Auffret, S.; Boulle, O.; Gaudin, G.; Chshiev, M.; Baraduc, C.; Bea, H. Large-Voltage Tuning of Dzyaloshinskii-Moriya Interactions: A Route toward Dynamic Control of Skyrmion Chirality. *Nano Lett.* **2018,** 18, (8), 4871-4877.

36. Diez, L. H.; Liu, Y. T.; Gilbert, D. A.; Belmeguenai, M.; Vogel, J.; Pizzini, S.; Martinez, E.; Lamperti, A.; Mohammedi, J. B.; Laborieux, A.; Roussigne, Y.; Grutter, A. J.; Arenholtz, E.; Quarterman, P.; Maranville, B.; Ono, S.; El Hadri, M. S.; Tolley, R.; Fullerton, E. E.; Sanchez-Tejerina, L.; Stashkevich, A.; Cherif, S. M.; Kent, A. D.; Querlioz, D.; Langer, J.; Ocker, B.; Ravelosona, D. Nonvolatile Ionic Modification of the Dzyaloshinskii-Moriya Interaction. *Phys. Rev. Appl.* **2019,** 12, (3), 034005.

37. Chen, G.; Mascaraque, A.; Jia, H. Y.; Zimmermann, B.; Robertson, M.; Lo Conte, R.; Hoffmann, M.; Barrio, M. A. G.; Ding, H. F.; Wiesendanger, R.; Michel, E. G.; Blügel, S.; Schmid, A. K.; Liu, K. Large Dzyaloshinskii-Moriya interaction induced by chemisorbed oxygen on a ferromagnet surface. *Sci. Adv.* **2020,** 6, (33), eaba4924.

38. Chen, G.; Robertson, M.; Hoffmann, M.; Ophus, C.; Cauduro, A. L. F.; Lo Conte, R.; Ding, H. F.; Wiesendanger, R.; Blugel, S.; Schmid, A. K.; Liu, K. Observation of Hydrogen-Induced Dzyaloshinskii-Moriya Interaction and Reversible Switching of Magnetic Chirality. *Phys. Rev. X* **2021,** 11, (2), 021015.

39. Nembach, H. T.; Jue, E.; Poetzger, K.; Fassbender, J.; Silva, T. J.; Shaw, J. M. Tuning of the Dzyaloshinskii-Moriya interaction by He+ ion irradiation. *J. Appl. Phys.* **2022,** 131, (14), 143901.

40. Vedmedenko, E. Y.; Udvardi, L.; Weinberger, P.; Wiesendanger, R. Chiral magnetic ordering in two-dimensional ferromagnets with competing Dzyaloshinsky-Moriya interactions. *Phys. Rev. B* **2007,** 75, (10), 104431.

41. Meckler, S.; Mikuszeit, N.; Pressler, A.; Vedmedenko, E. Y.; Pietzsch, O.; Wiesendanger, R. Real-space observation of a right-rotating inhomogeneous cycloidal spin spiral by spin-polarized scanning tunneling microscopy in a triple axes vector magnet. *Phys. Rev. Lett.* **2009,** 103, (15), 157201.

42. Wiesendanger, R. Nanoscale magnetic skyrmions in metallic films and multilayers: a new twist for spintronics. *Nat. Rev. Mater.* **2016,** 1, (7), 16044.

43. Chen, G.; Zhu, J.; Quesada, A.; Li, J.; N'Diaye, A. T.; Huo, Y.; Ma, T. P.; Chen, Y.; Kwon, H. Y.; Won, C.; Qiu, Z. Q.; Schmid, A. K.; Wu, Y. Z. Novel chiral magnetic domain wall structure in Fe/Ni/Cu(001) films. *Phys. Rev. Lett.* **2013,** 110, (17), 177204.

44. Yang, H.; Chen, G.; Cotta, A. A. C.; N'Diaye, A. T.; Nikolaev, S. A.; Soares, E. A.; Macedo, W. A. A.; Liu, K.; Schmid, A. K.; Fert, A.; Chshiev, M. Significant Dzyaloshinskii-Moriya interaction at graphene-ferromagnet interfaces due to the Rashba effect. *Nat. Mater.* **2018,** 17, (7), 605-609.

45. Sander, D. The correlation between mechanical stress and magnetic anisotropy in ultrathin films. *Rep. Prog. Phys.* **1999,** 62, (5), 809-858.

46. Rohart, S.; Thiaville, A. Skyrmion confinement in ultrathin film nanostructures in the presence of Dzyaloshinskii-Moriya interaction. *Phys. Rev. B* **2013,** 88, (18), 184422.

47. Braun, H.-B. Topological effects in nanomagnetism: from superparamagnetism to chiral quantum

solitons. *Adv. Phys.* **2012,** 61, (1), 1-116.

48. Bhattacharya, D.; Razavi, S. A.; Wu, H.; Dai, B.; Wang, K. L.; Atulasimha, J. Creation and annihilation of non-volatile fixed magnetic skyrmions using voltage control of magnetic anisotropy. *Nat. Electron.* **2020,** 3, (9), 539-545.

49. Chen, G.; Ophus, C.; Quintana, A.; Kwon, H.; Won, C.; Ding, H.; Wu, Y.; Schmid, A. K.; Liu, K. Reversible writing/deleting of magnetic skyrmions through hydrogen adsorption/desorption. *Nat. Commun.* **2022,** 13, (1), 1350.

50. Camosi, L.; Rougemaille, N.; Fruchart, O.; Vogel, J.; Rohart, S. Micromagnetics of antiskyrmions in ultrathin films. *Phys. Rev. B* **2018,** 97, (13), 134404.

51. Emori, S.; Bauer, U.; Ahn, S. M.; Martinez, E.; Beach, G. S. Current-driven dynamics of chiral ferromagnetic domain walls. *Nat. Mater.* **2013,** 12, (7), 611-616.

52. Berruto, G.; Madan, I.; Murooka, Y.; Vanacore, G. M.; Pomarico, E.; Rajeswari, J.; Lamb, R.; Huang, P.; Kruchkov, A. J.; Togawa, Y.; LaGrange, T.; McGrouther, D.; Ronnow, H. M.; Carbone, F. Laser-Induced Skyrmion Writing and Erasing in an Ultrafast Cryo-Lorentz Transmission Electron Microscope. *Phys. Rev. Lett.* **2018,** 120, (11), 117201.

53. Wang, Z. D.; Guo, M. H.; Zhou, H. A.; Zhao, L.; Xu, T.; Tomasello, R.; Bai, H.; Dong, Y. Q.; Je, S. G.; Chao, W. L.; Han, H. S.; Lee, S.; Lee, K. S.; Yao, Y. Y.; Han, W.; Song, C.; Wu, H. Q.; Carpentieri, M.; Finocchio, G.; Im, M. Y.; Lin, S. Z.; Jiang, W. J. Thermal generation, manipulation and thermoelectric detection of skyrmions. *Nat. Electron.* **2020,** 3, (11), 672-679.

54. Weller, D.; Parker, G.; Mosendz, O.; Lyberatos, A.; Mitin, D.; Safonova, N. Y.; Albrecht, M.; D., W.; A., M. Review Article: FePt heat assisted magnetic recording media. *J. Vac. Sci. Technol. B* **2016,** 34, (6), 060801.